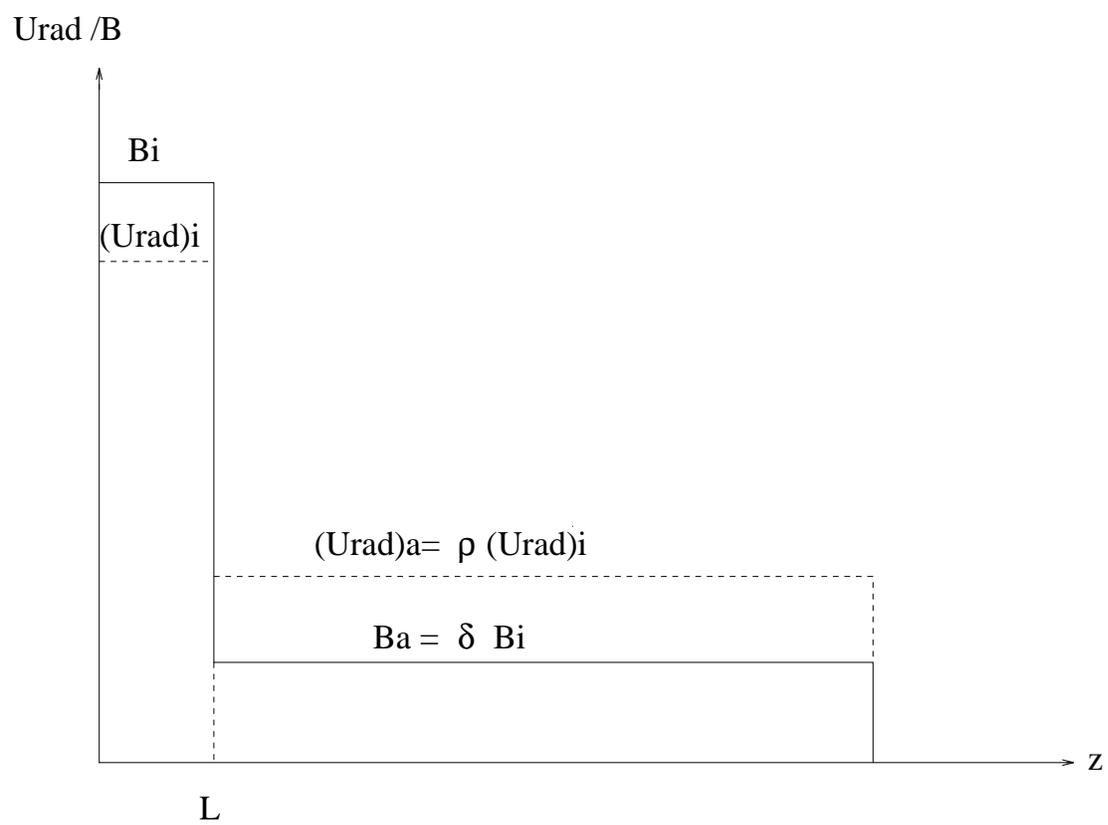

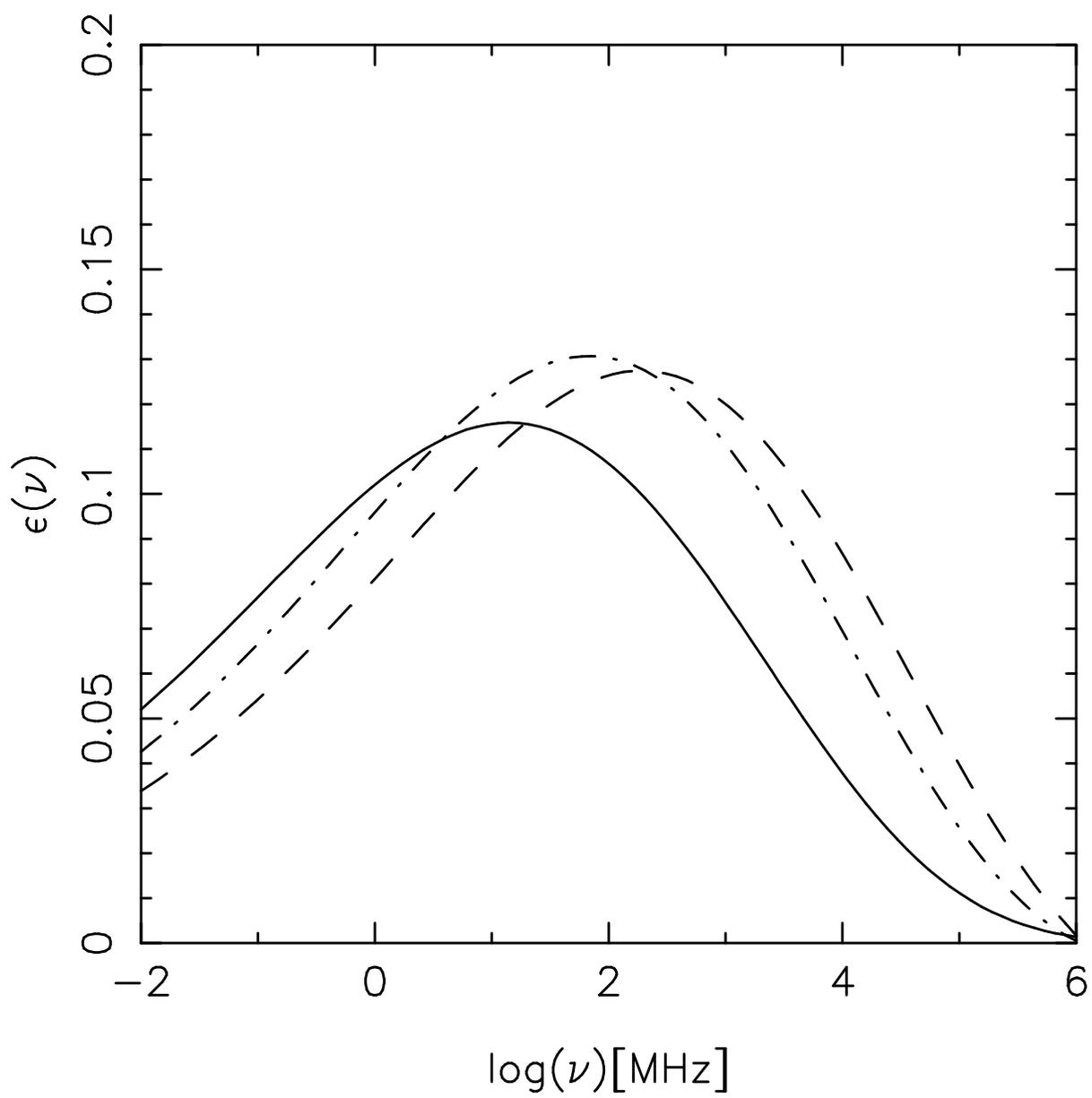

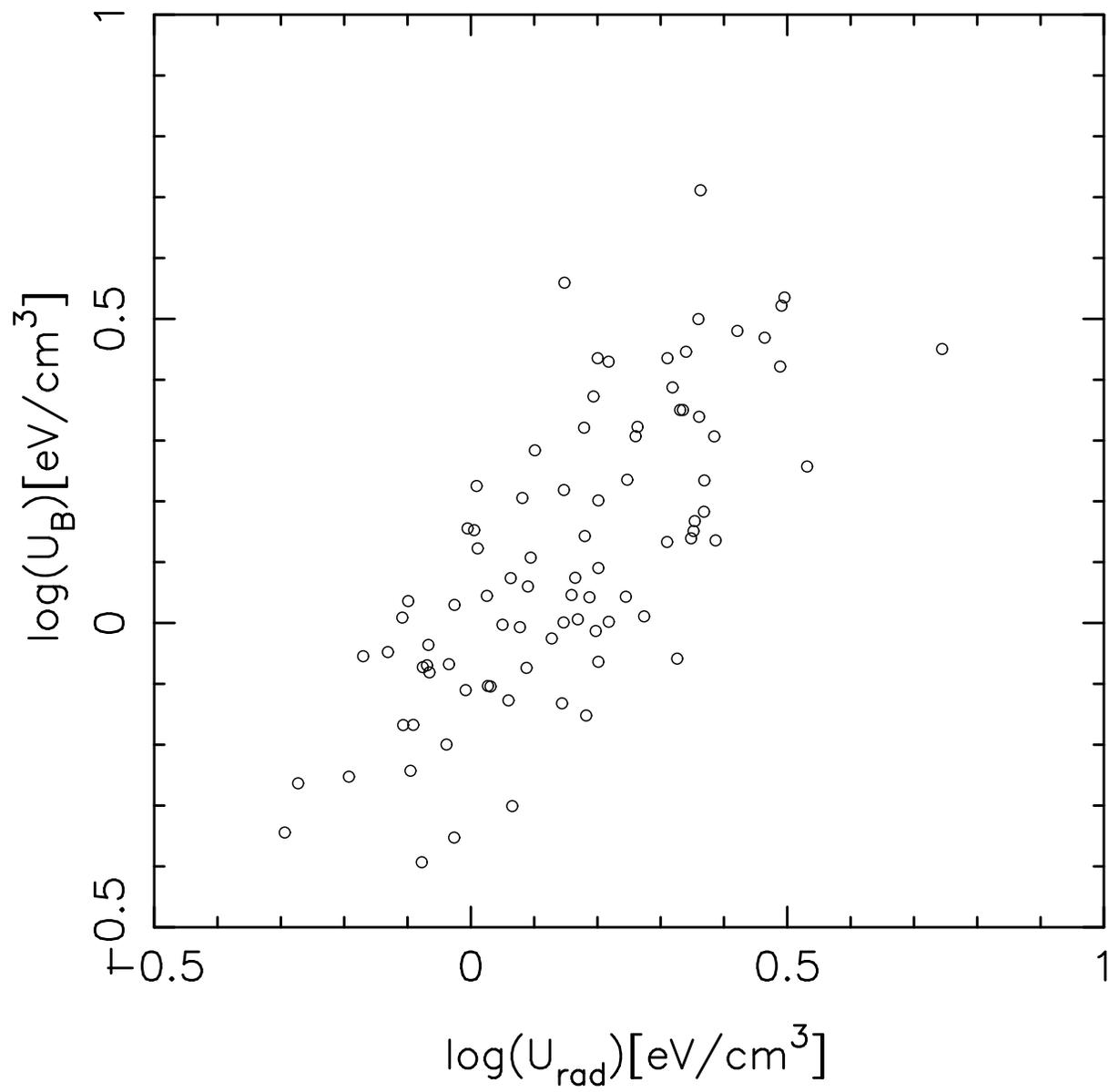

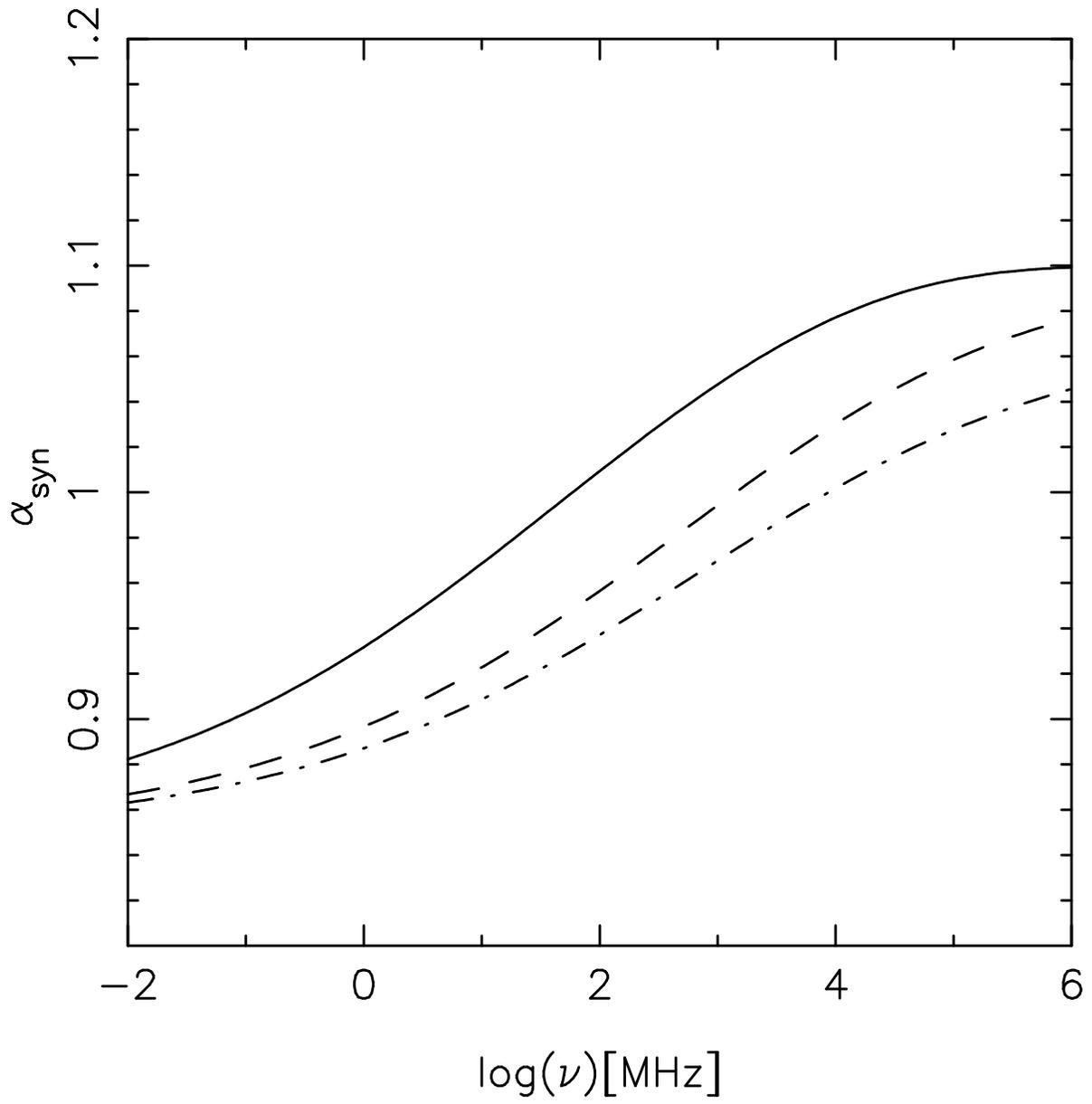

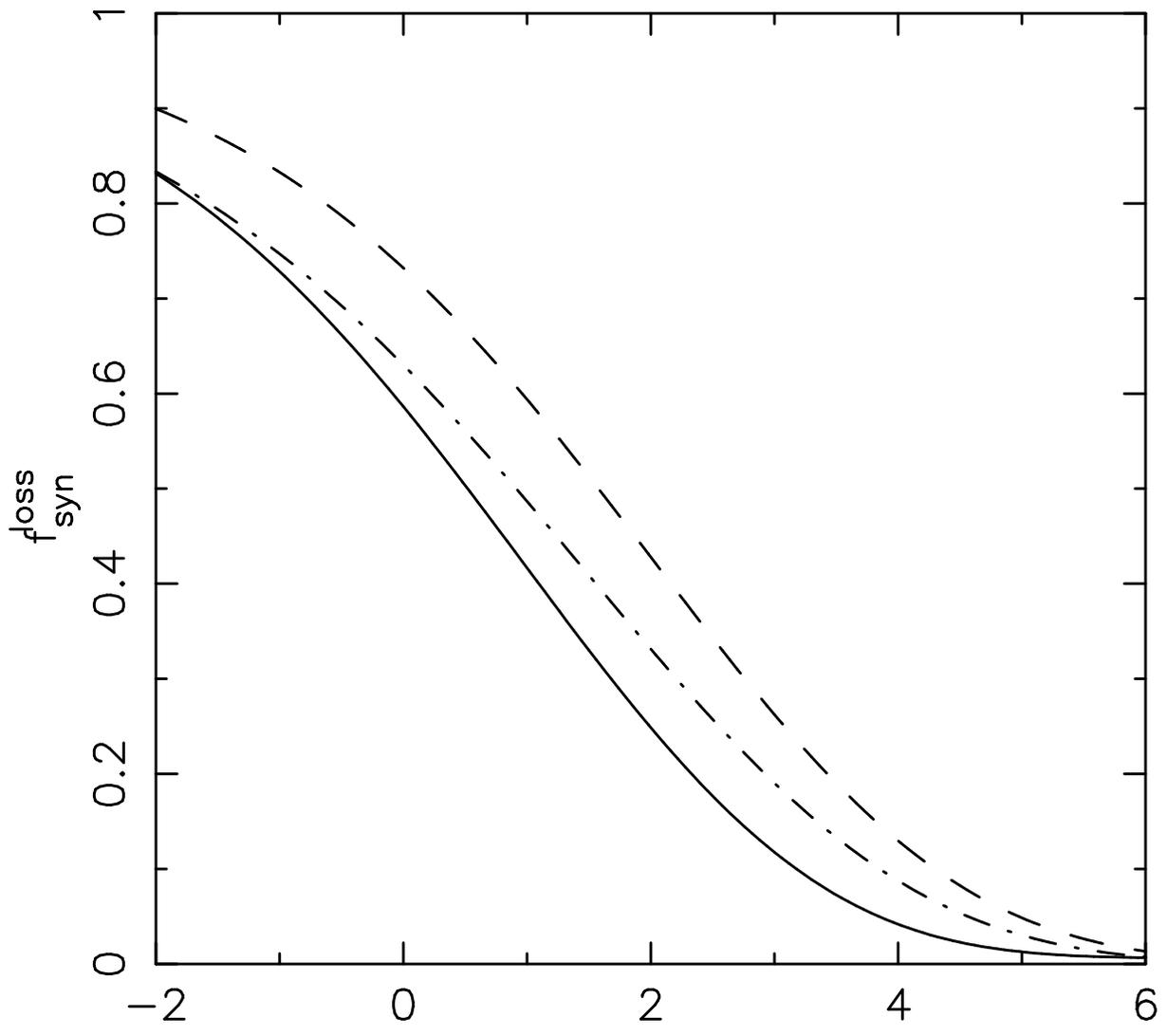

# A quantitative model of the FIR/radio correlation for normal late-type galaxies

U. Lisenfeld*, H.J. Völk, and C. Xu

Max-Planck-Institut für Kernphysik, Postfach 10 35 80, D-69029 Heidelberg, Germany

**Abstract.** This paper investigates the physical reasons for the existence, the tightness and the near universality of the FIR/radio correlation for late-type field galaxies whose emission is not dominated by an active nucleus. We develop theoretical models for the radio and far-infrared (FIR) emission of such normal galaxies and study the influence of their main parameters on the ratio of the two emissions. In addition, data are used from a sample of 114 late-type galaxies which allow an estimate of the mean energy densities of the radiation field and the magnetic field, the latter crudely calculated from the synchrotron luminosity using the minimum energy condition, and of the dust opacity. These data reveal, for the first time, a reasonably good, linear correlation between the energy density of the radiation field and the energy density of the magnetic field. Interestingly this implies that the two most important energy loss rates for electrons, synchrotron and Inverse Compton losses, are proportional to each other. As a consequence the radio synchrotron emission is proportional to the total flux of radiative energy loss from the nonthermal Cosmic Ray electrons of a given energy. Furthermore we find that on average the galaxies are marginally optically thick for the non-ionizing UV light. Including their extended magnetic halos, galaxies are also found to be on average marginally optically thick regarding the radiative energy losses of the radio synchrotron emitting, non-thermal Cosmic Ray electrons. Exceptions may be galaxies with a very low (compared to our Galaxy) present star formation rate. Combining these semi-empirical results with the theoretical emission models we show first of all that in the optically thick case the linear correlation between the energy densities of the radiation field and the magnetic field is a necessary condition for the existence of the observed FIR/radio correlation. Secondly we show that the individual dispersion of the FIR to radio continuum ratio caused by uncertainties of any single one of the parameters considered is significantly less than the empirical dispersion of the FIR/radio correlation. The radio and the FIR emission are mainly determined by their sources, i.e. massive Supernova precursor stars, because galaxies act in a reasonable approximation as calorimeters for the stellar UV radiation and for the energy flux of the Cosmic Ray electrons produced in their disks. This constitutes a very general theoretical explanation for the FIR/radio correlation. In principle one can now go a significant step further and use this observed correlation plus the theory and the parameters of one galaxy as a normalization to deduce in turn the mean magnetic field strength of other normal galaxies simply from their total mean radiation energy density.

## 1. Introduction

The nearly universal correlation between the integrated far-infrared (FIR) and radio continuum emission of late-type field galaxies (see Helou 1991 for a review of the observations) is best considered in the framework of a common dependence of the FIR and the radio continuum emission on the formation of massive stars: (i) the FIR emission is mainly due to dust heated by the ionizing and non-ionizing UV radiation from these stars, (ii) their ionizing radiation is the source of the thermal radio emission, (iii) the Supernova (SN) explosions which occur at the end of their lives accelerate the Cosmic Ray (CR) electrons responsible for the nonthermal radio emission (see Condon 1992, and Völk & Xu 1994, for reviews). However, the actual existence of a correlation and especially its extreme tightness ($\sim 0.2$ in the logarithm, Wunderlich et al. 1987) which is the strongest among various correlations found for many different radiations has been difficult to understand, given the very different radiation mechanisms of the FIR and the radio continuum.

So far the problem has been addressed, in a qualitative manner, in two different asymptotic theories assuming exactly the opposite extreme conditions. In an 'optically-thick' picture, the so-called "calorimeter theory" (Völk 1989; Xu 1990; Lisenfeld

*Send offprint requests to*: U. Lisenfeld
* Present address: Mullard Radio Astronomy Observatory, Cavendish Laboratory, Madingley Road, Cambridge CB3 OHE, Britain

1993; Völk and Xu 1994), it is argued that both the very existence and the tightness of the correlation may be explained by three plausible but far reaching properties of galaxies: First of all the main part of the relevant stellar radiation, presumably the UV radiation from massive stars, is absorbed by the dust grains *within* a galaxy. Secondly, also the energetic CR electrons that are produced by eventual SN explosions of these stars lose their energy mainly *within* the galaxy and its halo, predominantly by Inverse Compton collisions with the photons of the interstellar radiation field as well as by synchrotron radiation in the magnetic field. In this case the galaxy as a whole acts like a calorimeter for its own primary photon and electron emission, independent of the details of production and transport of these two carriers of energy. Finally, the energy density of the interstellar radiation field, $U_{rad}$, is in a *constant ratio* with the magnetic field energy density, $U_B$. This means that the radio synchrotron emission is proportional to the total (radiative) energy loss rate, and that both FIR and nonthermal radio emission are essentially proportional to the formation rate of the massive stars with but a weak dependence on all other parameters. The result is a tight, roughly linear correlation. When the contribution from the old stars to the heating of dust is taken into account, the weak nonlinearity of the correlation ($P_{1.49\text{GHz}} \propto L_{\text{FIR}}^{1.3}$) can also be explained satisfactorily (Xu et al. 1994).

Alternatively, in the 'optically-thin' picture (Helou & Bicay 1993; Chi & Wolfendale 1990), it is assumed that most of the radiation which is responsible for the heating of dust, and most of the CR electrons which produce synchrotron radiation, escape from a typical galaxy disk, i.e. disk galaxies are optically thin for both the primary photons and CR electrons. If this is indeed the case for the majority of late-type galaxies, it is argued (Helou & Bicay 1993) that the FIR/radio correlation indicates a proportionality between the two escape rates, namely $\tau_0 \propto t_{esc}/t_{sync}$ where $\tau_0$ is the optical depth of the disk for the primary photons, and $t_{esc}$ and $t_{sync}$ the escape time scale and synchrotron loss time scale of CR electrons. The nonlinearity of the FIR/radio correlation is then interpreted as a consequence of the dependence of the CR electron escape rate on the FIR luminosity (Chi & Wolfendale 1990).

In this paper we shall study the FIR/radio correlation with a quantitative perspective and scrutinize these asymptotic theories. Theoretical models for both the radio continuum and the FIR emission will be developed. We will show explicitly that a large number of parameters are involved. This demonstrates the significance of the correlation. For the major parameters determining the emissions, e.g. the optical depth, the magnetic field strength, the ratio between the energy densities of the radiation field and the magnetic field, etc., we will avoid a priori assumptions but rather constrain their mean values and dispersions using observational data from a sample of 114 nearby spiral galaxies and other data from the literature. It is clear that such a correlation can only exist for normal field galaxies, outside the core of galaxy clusters, because the action of an active nucleus (AGNs) or the intense interaction with other galaxies would introduce effects independent of the star formation and destruction processes. Hence we will not consider galaxies with an AGN as well as clustered galaxies.

In Sect. 2 we present the models for the FIR and the radio emission. In Sect. 3 we discuss observational constraints for some of the parameters, especially the ratio $U_{rad}/U_B$. In Sect. 4 the sources of the dispersion of the FIR/radio ratio are evaluated, whereas Sect. 5 presents our results concerning the question of the optical depth of galaxies. Sect. 6 contains the discussion. In Sect. 7 the main conclusions are summarized.

## 2. Models of the radio continuum and the FIR emissions

Many parameters, most of which do not depend directly on the formation rate of massive stars, are involved in the radiation processes of the radio continuum and of the FIR emission of galaxies. In order to study their influence on the FIR/radio flux ratio, we give in the following a detailed theoretical description of the mechanisms of the two emissions, and evaluate the importance of individual parameters for the corresponding process.

### 2.1. Dust heating and FIR emission

The FIR radiation of normal galaxies is completely dominated by the thermal emission of dust heated by stellar radiation. The FIR luminosity (in the wavelength range of 40–120$\mu m$) of such a galaxy can be derived from the following formula:

$$L_{\text{FIR}} = \int_{m1}^{m2} \zeta(m)\Phi(m)\,dm \int_0^{t'(m)} r(t)\,dt \quad (1)$$
$$\int P_\lambda(m) f_\lambda(m)\,d\lambda$$

with $m_1 = 1 M_\odot$ and $m_2 = 100 M_\odot$ specifying the mass range of stars which are responsible for the heating of dust (we ignore stars less massive than the Sun because they contribute little to the dust heating, and take an upper cut-off of $100 M_\odot$ for the Initial Mass Function (IMF) $\Phi(m)$); $t'(m)$ is the main sequence life time of a star of mass m, $f_\lambda(m)$ the radiation spectrum of a star, and $\zeta(m)$ is the fraction of the absorbed radiation from a certain star to be re-radiated within the wavelength range 40–120$\mu m$. From equation (1), the FIR luminosity depends in addition on the star formation rate and its time dependence $r(t)$, where $t$ is the looking back time, as well as on the probability, $P_\lambda(m)$, of radiation at wavelength $\lambda$, radiated by the star of mass m, to be absorbed by dust.

In order to parameterize these dependences, we make the following assumptions:
(1) The IMF is a power law between $m_1 = 1$ $M_\odot$ and $m_2 = 100$ $M_\odot$, with power index k :

$$\Phi(m) = \Phi_0 \; m^{-k} \quad (1 \leq m/M_\odot \leq 100). \quad (2)$$

(2) The stars in the mass range 1 — 100 $M_\odot$ can be divided into 3 populations: i) the massive ionizing stars of

$20 < m \leq 100 M_\odot$, with a life time $t'(m) < 10^7$ years, so that $\int_0^{t'(m)} r(t)\, dt \simeq r_0 \times t'(m)$, where $r_0 = r(t=0)$ is the current star formation rate, ii) the intermediate massive stars of $5 < m \leq 20 M_\odot$ which are mostly responsible for the non-ionizing UV radiation of a galaxy, with a characteristic life time $t'(m) = 10^8$ years, iii) the low mass stars of $1 < m \leq 5 M_\odot$ which are mainly responsible for the blue radiation of a galaxy, with a characteristic life time of $t'(m) = 3 \cdot 10^9$ years. For populations ii) and iii) the time integration of the star-formation rate can be approximated by $r_8 \times t'(m)$ and $r_9 \times t'(m)$, where $r_8$ and $r_9$ are the star formation rate averaged over the last $10^8$ years and $3 \cdot 10^9$ years, respectively (Xu et al. 1994).

(3) The massive ionizing stars heat the dust locally, i.e. they are the sources of the HII-region-associated FIR emission. The absorption probability of their radiation is taken as a constant (Xu et al. 1994):

$$P_\lambda(m) = P_{HII} = 0.6 \qquad (m > 20 M_\odot) \qquad (3)$$

Using the model of Désert et al. (1990) for the dust heated by an O5 star (dilution factor X=0.01), we estimate that the fraction of the HII-region-associated dust emission through the window 40–120$\mu m$ amounts to

$$\zeta(m) = \zeta_{HII} = 0.5 \qquad (m > 20 M_\odot). \qquad (4)$$

(4) Dust heating due to intermediate massive stars and old stars occurs through the diffuse and approximately uniform interstellar radiation field which heats the diffuse dust in the disk. The resulting FIR emission is the so-called Cirrus. We take the corresponding $\zeta$ value in all galaxies as that of the cirrus emission in the solar neighbourhood (Désert et al. 1990), which is 30 %:

$$\zeta(m) = \zeta_{diff} = 0.3 \qquad (m \leq 20 M_\odot). \qquad (5)$$

(5) The radiative transfer for the non-ionizing emission of the intermediate massive stars and the old stars is modeled using an infinite parallel-slab geometry, assuming that stars and dust are uniformly mixed. According to Xu and De Zotti (1989) the absorption probability is then estimated from the optical depth of galactic disks:

$$P_\lambda(m) = \frac{\tau_\lambda'^{0.9}}{0.6 + \tau_\lambda'^{0.9}} \qquad (m \leq 20 M_\odot) , \qquad (6)$$

where $\tau_\lambda' = \beta \tau_\lambda$ is the so called 'effective optical depth' of the disk which takes into account the effect of scattering, and $1 - a_\lambda < \beta < 1$, where $a_\lambda \sim 0.6$ is the albedo (Mathis et al. 1983). We take $\beta = 0.7$. For the sake of simplicity, we make the further approximations:

$$P_\lambda(m) = \begin{cases} P_{UV} = \frac{\tau_{UV}'^{0.9}}{0.6 + \tau_{UV}'^{0.9}} & (5 < m \leq 20 M_\odot) \\ P_B = \frac{\tau_B'^{0.9}}{0.6 + \tau_B'^{0.9}} & (1 < m \leq 5 M_\odot) \end{cases} \qquad (7)$$

where $\tau_{UV}'$ is the effective UV optical depth at 2000Å, and $\tau_B'$ is the effective blue optical depth at 4300Å. The two optical depths are related to each other through the adopted extinction curve (Savage and Mathis 1979). This gives $\tau_B' = 0.47\, \tau_{UV}'$.

With the above assumptions and simplifications, equation (1) can be rewritten as:

$$\begin{aligned} L_{\text{FIR}} =& r_0\, P_{HII}\, \zeta_{HII}\, L_i(k) + r_8\, P_{UV}(\tau_{UV})\, \zeta_{diff}\, L_{ii}(k) \\ & + r_9\, P_B(\tau_B)\, \zeta_{diff}\, L_{iii}(k) \end{aligned} \qquad (8)$$

where

$$L_i(k) = \int_{20 M_\odot}^{100 M_\odot} \Phi_0 m^{-k}\, t'(m) L(m)\, dm \qquad (9)$$

is the total luminosity of massive ionizing stars for $r_0 = 1$, and $L(m) = \int f_\lambda(m)\, d\lambda$ is the total luminosity of a star of mass m. Since the product of the life time and the luminosity for stars of a given mass is well determined from the theory of stellar evolution, $L_i$ is only a function of the parameter k which specifies the IMF. The same is true for the similar variables $L_{ii}$ (for the intermediate massive stars – between 5 and 20 $M_\odot$) and $L_{iii}$ (for the old stars – between 1 and 5 $M_\odot$). According to Xu et al. (1994) we will assume for the total radiation emitted during a stellar main-sequence life-time:

$$t'L = \begin{cases} 10^{9.95} m\ L_\odot yr & (1 M_\odot \leq m < 5 M_\odot) \\ 10^{9.6} m^{3/2}\ L_\odot yr & (5 M_\odot \leq m < 100 M_\odot) \end{cases} \qquad (10)$$

In Lisenfeld (1993) a more detailed model for the dust heating has been considered in which the dust absorption of non-ionizing radiation ($\lambda > 912$Å) and the dust absorption of ionizing radiation in HII regions are explicitly taken into account. The fraction of stellar radiation absorbed by dust could therefore be calculated without the need of simplifying assumptions like Eqs. (3) and (7). However, the results of Lisenfeld (1993) support the validity of these assumptions: Taking as an example $\tau_{UV} = 1$, Eqs. (3) and (7)) predict that in the mass range of 1-5 $M_\odot$ (respectively 5-20 $M_\odot$, 20-100 $M_\odot$) a fraction of 37 % (54 %, 60%) of the radiation of a star is absorbed. The more detailed calculations yield for the same mass ranges 25 -40 % (40-60 %, 60-65%), where the smaller numbers of theses ranges refer to the low mass ends of the mass intervals, and the larger numbers to the high mass ends. For different dust opacities the agreement is equally good.

### 2.2. Radio continuum emission

#### 2.2.1. Synchrotron emission of CR electrons

In a one-dimensional, steady state diffusion model, the diffusion equation describing the production, diffusion and energy losses of electrons reads as follows (e.g. Berezinsky et al. 1990):

$$D(E)\frac{\partial^2 f(E,z)}{\partial z^2} + \frac{\partial}{\partial E}\left(b(E) f(E,z)\right) = -Q(E,z) , \qquad (11)$$

where $z$ is the height above the galactic midplane, and $f(E,z)$ denotes the phase space density of the electrons, i.e. $f(E,z)$

$$D(E) = D_0 \left(\frac{E}{1\text{GeV}}\right)^\mu \tag{12}$$

is the energy dependent diffusion coefficient, and $b(E, z)$ describes the electron energy losses in the energy range considered here ($E \gtrsim 1$ GeV). These are dominantly Inverse Compton and synchrotron losses:

$$b(E, z) = -\left.\frac{dE}{dt}\right|_{syn+IC} = \frac{4}{3}\sigma_T c \left(\frac{E}{m_e c^2}\right)^2 (U_{\text{rad}} + U_B). \tag{13}$$

Here, $\sigma_T$ is the Thompson scattering cross section, $c$ is the speed of light; $U_B$ denotes the energy density of the magnetic field and $U_{\text{rad}}$ the energy density of the radiation field below the Klein-Nishina limit (e.g. Longair 1992). For the electron energies considered here ($E \gtrsim 1$ GeV), this limit for the photon energies lies in the X-ray domain. Thus, to a very good approximation $U_{\text{rad}}$ represents the energy density of the entire radiation field. In order to simplify Eq. (11), we make the following assumptions:

1) The CR electrons are produced only by Supernova Remnants (SNRs) from massive stars which are situated in the galactic midplane at z=0, and the production rate is proportional to the SN rate. The source function has a power-law spectrum (e.g. Völk et al. 1988):

$$Q(E, z) = \delta(z) \left(\frac{E}{m_e c^2}\right)^{-x} \nu_{\text{SN}} q_{\text{SN}}, \tag{14}$$

where $\nu_{\text{SN}}$ is the SN rate, $q_{\text{SN}}$ is the average of the number of relativistic electrons that are produced per energy interval dE by a SN, and $x \geq 2$ denotes the spectral index. We have therefore neglected other possible CR electron production mechanisms, e.g. shocks associated with type I SN from (old) White Dwarfs in accreting binary systems (Xu. et al. 1994), stellar winds (Cesarsky and Montmerle 1983), spiral arms (Duric 1986), and pulsars (e.g. Völk 1992). Assuming that stars more massive than 5$M_\odot$ are precursors of the SN considered, $\nu_{SN}$ is determined by the adopted IMF (i.e. the parameter k) and by $r_8$, the star formation rate averaged over the last $10^8$ years, (the life time of intermediate massive stars which considerably outnumber the massive stars):

$$\nu_{\text{SN}} = \int_{5M_\odot}^{100M_\odot} \Phi_0 m^{-k} s(t = t'(m)) \, dm \tag{15}$$
$$\simeq r_8 N_{SN}(k)$$

where $N_{\text{SN}}(k) \simeq \Phi_0 \times 5^{-(k-1)}/(k-1)$ is the number of SN when $r(t) = 1$ yr$^{-1}$.

2) The energy densities $U_{\text{rad}}$ of the radiation field and $U_B$ of the magnetic field (and therefore also the field strength B) are step functions of the height z over the galactic plane:

$$B = \begin{cases} B_i & (|z| \leq L) \\ \delta B_i & (|z| > L) \end{cases}$$

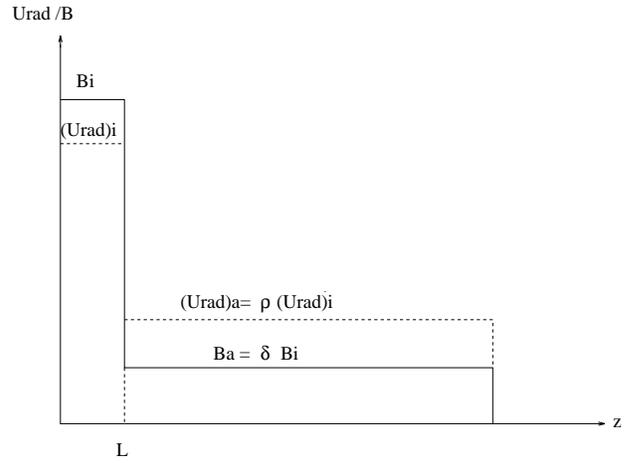

**Fig. 1.** Geometry of the magnetic field $B$ and the energy density of the radiation field $U_{\text{rad}}$ in the galactic halo: The magnetic field $B_a$ and the energy density of the radiation field $(U_{\text{rad}})_a$ in the outer halo ($z > L$) are by a factor $\delta$ and $\rho$, respectively, lower than the values in the inner halo, $B_i$ and $(U_{\text{rad}})_i$.

$$U_{rad} = \begin{cases} (U_{rad})_i & (|z| \leq L) \\ \rho(U_{rad})_i & (|z| > L), \end{cases} \tag{16}$$

where L ($\sim 1$ kpc) is a free parameter to separate the so defined outer halo from the inner halo, and $\delta < 1$ as well as $\rho < 1$ (Fig.1). This is a first order approximation to describe the actual spatial decrease of the two fields with distance from the galactic disk.

3) A free escape boundary exists at $|z| = z_h$, the outer boundary of the halo:

$$f(E, |z_h|) = 0 \tag{17}$$

(e.g. Ginzburg & Syrovatskii, 1964, p. 300). An approximate solution of Eq. (11) for a uniform halo (i.e. $\delta = \rho = 1$) is given by:

$$f(E, z) = \frac{\nu_{SN} q_{SN}}{2D_0} \left(\frac{E}{m_e c^2}\right)^{-x} \left(\frac{E}{1GeV}\right)^{-\mu}$$
$$\left(\frac{{}_1F_1(p+\frac{1}{2}, \frac{3}{2}, s_h)}{{}_1F_1(p, \frac{1}{2}, s_h)} {}_1F_1(p, \frac{1}{2}, s) \cdot |z_h| \right. \tag{18}$$
$$\left. - {}_1F_1(p+\frac{1}{2}, \frac{3}{2}, s) \cdot |z| \right)$$

with

$$p = -\left(\frac{2x + \mu - 3}{2(1-\mu)}\right), \tag{19}$$

${}_1F_1(a, b, x)$ denoting the confluent hypergeometric function (or Kummer function) (e.g. Mathews and Walker, 1970), and

$$s = -\frac{c_1(U_{\text{rad}} + U_B)(1-\mu)}{4D_0} z^2 \left(\frac{E}{m_e c^2}\right) \left(\frac{E}{1GeV}\right)^{-\mu} \tag{20}$$

$$s_h = -\frac{c_1(U_{\rm rad}+U_{\rm B})(1-\mu)}{4D_0} z_h^2 \left(\frac{E}{m_e c^2}\right)\left(\frac{E}{1GeV}\right)^{-\mu} \quad (21)$$

where

$$c_1 = \frac{4}{3}\frac{\sigma_T c}{m_e c^2}. \quad (22)$$

These variables are proportional to the ratio of the diffusion time scale $\tau_{diff}$ to the energy loss time scale $\tau_{loss}$:

$$s = -\left(\frac{1-\mu}{4}\right)\frac{\tau_{diff}}{\tau_{loss}} \quad (23)$$

with

$$\tau_{diff} = \frac{z^2}{D_0(E/1GeV)^\mu} \quad (24)$$

and

$$\tau_{loss} = \int_E^\infty \left(\frac{dE}{dt}\right)^{-1} dE = \frac{1}{c_1(E/m_e c^2)(U_{\rm rad}+U_{\rm B})} \quad (25)$$

For a nonuniform halo, a similar approximate solution can be found with an analogous substitution by solving Eq.(11) separately for the inner and outer part of the halo. The solutions are matched at $z = L$ such that both the density $f(E,z)$ and the flux $D(E)(\partial f(E,z)/\partial z)$ are continuous (Lisenfeld 1993). The accuracy of these solutions is discussed in Appendix A.

The total synchrotron emission $P_{syn}(\nu)$ of a galaxy is calculated as:

$$P_{syn}(\nu) = \int_{-z_h}^{z_h} dz f(E,z)\frac{dE}{dt}\bigg|_{syn}\frac{dE}{d\nu}. \quad (26)$$

Here we use the simplifying assumption that an electron emits its synchrotron radiation completely at the maximum frequency $\nu$ of its synchrotron spectrum:

$$\nu = \left(\frac{E}{m_e c^2}\right)^2 \nu_G \ ,$$

where $\nu_G = eB/m_e c^2$ is the gyrofrequency.

We assume a constant spectral index of the injection spectrum $x = 2.2$ (Völk et al. 1988) although theoretically x = 2.1 (Berezkho et al. 1993) is an equally plausible value (which then requires $\mu \simeq 0.6$ to establish agreement with CR observations in our Galaxy).

We further assume that each SN (mainly type II and Ib) is a "standard candle", i.e. they all produce the same amount and the same spectrum of relativistic electrons (Markiewicz, Drury & Völk 1990). Although this might not hold precisely for individual SN, it is probably a good assumption when averaged over a whole galaxy since at any given epoch $\sim 1000$ SNRs act as CR sources taking our Galaxy as a typical case (with an active life time $\sim 10^4$ to $10^5$ yrs for a SNR and $\nu_{\rm SN} \sim 1/30$yr). Under this assumption, the parameter $q_{SN}$ is a constant. Its value can be deduced from the data of the Galaxy: The intensity of synchrotron radiation of the Galaxy at 20cm is $P_{syn} = 2.7\ 10^{21}$ W/Hz (extrapolating the results of Beuermann et al. 1985 and assuming a nonthermal spectral index of 0.9), the SN rate is, within a factor of about 2, $\nu_{SN} = 0.04$ yr$^{-1}$ (van den Bergh and Tammann, 1991), the strength of magnetic field is $B = 5\mu G$ (Sofue, Fujimoto & Wielebinski 1986) and $D_0 = 10^{29}$ cm$^2$/sec, $z_h = 10$kpc (Berezinsky et al. 1990). From these data we derive the value of $q_{SN} = 1.6\ 10^{48}$ eV$^{-1}$. The value of $q_{SN}$ is, mainly because of the uncertainty in the Galactic SN-rate, at least uncertain to within a factor of 2 (0.3 in the logarithm).

We can therefore parameterize the total synchrotron emission of a galaxy in the following symbolic form:

$$P_{syn}(\nu) = q_{SN}\nu_{SN} \\ \Pi_\nu(x,\mu,(U_{\rm rad}/U_{\rm B})_i, B_i, D_0, z_h, \delta, \rho, L), \quad (27)$$

where $\Pi_\nu$ is a nonlinear function of the nine parameters indicated.

It is instructive to consider $P_{syn}$ in the asymptotic limits $|s_h| \gg 1$ and $|s_h| \ll 1$. This requires the asymptotic expansions of the confluent hypergeometric function:

$$_1F_1(a,b,x) \approx \begin{cases} 1 & \text{if } x \ll 1 \\ \frac{\Gamma(b)}{\Gamma(b-a)}x^{-a} & \text{if } x \gg 1 \end{cases} \quad (28)$$

where $\Gamma(x)$ denotes the gamma function (e.g. Mathews & Walker, 1970).

For a homogeneous halo we obtain for $|s_h| \ll 1$:

$$P_{syn}(\nu) = \frac{\nu_{SN} q_{SN}}{2D_0} c_1 \frac{U_{\rm B}}{\nu_G}(m_e c^2)^2 z_h^2 \left(\frac{\nu}{\nu_G}\right)^{(\frac{1-x-\mu}{2})} \\ \propto \nu^{(\frac{1-x-\mu}{2})} B^{(\frac{1+x+\mu}{2})}\frac{z_h^2}{D_0} \quad (29)$$

$|s_h(E)| \ll 1$, means that the electrons have not suffered any substantial energy losses when they reach $z_h$. This is reflected in Eq.(29) by the fact that the synchrotron emission does not depend on the energy losses. The diffusion coefficient and the halo size $z_h$, however, are important parameters in this 'optically thin' limiting case.

In the opposite case, $|s_h| \gg 1$, the synchrotron emission is given by:

$$P_{syn} = \nu_{SN} q_{SN}\frac{(m_e c^2)^2}{4(x-1)\nu_G}\left(\frac{\nu}{\nu_G}\right)^{\frac{-x}{2}}\frac{1}{1+U_{\rm rad}/U_{\rm B}} \\ \propto \nu^{-\frac{x}{2}} B^{\frac{2x-2}{2}}\left(1+\frac{U_{\rm rad}}{U_{\rm B}}\right)^{-1}. \quad (30)$$

In this case of high frequency $\nu$, energy losses dominate the electron propagation. The electrons loose their energy down to E or below before they reach the outer boundary of the halo. The strength and energy dependence of the diffusion coefficient are not important for the total synchrotron emission (for a homogeneous halo) and therefore neither $D_0$ nor $\mu$ appears. This asymptotic case corresponds to the "calorimeter" theory considered by Völk (1989). It is characterized by a weak dependence on the magnetic field. In addition, $P_{syn}$ is inversely proportional to $(1 + U_{\rm rad}/U_{\rm B})$.

### 2.2.2. Thermal bremsstrahlung of ionized gas

Especially at high frequencies, thermal bremsstrahlung of ionized gas contributes to the radio emission as well. It is mainly determined by the number of Lyman continuum photons, $N_{\text{Lyc}}$ (in sec$^{-1}$), emitted by exciting stars:

$$P_{th}(\nu) = 2.52 \cdot 10^{-35} f_{ion} T_e^{0.45} \left( \frac{N_{\text{Lyc}}}{sec^{-1}} \right)$$
$$\left( \frac{\nu}{GHz} \right)^{-0.1} [W/Hz] \quad (31)$$

(e.g. Condon, 1992), where $f_{ion}$ is the fraction of Lyman continuum photons absorbed by the gas, $T_e$ is the electron temperature and $N_{\text{Lyc}}$ is a function of the IMF as well as of the present star formation rate $r_0$:

$$N_{\text{Lyc}} = \int_{20 M_\odot}^{100 M_\odot} \Phi_0 m^{-k} \tilde{n}_{Lyc}(m) s(t = t'(m)) \, dm \quad (32)$$
$$\simeq r_0 n_{Lyc}(k)$$

where $\tilde{n}_{Lyc}(m)$ is the production rate of Lyman continuum photons by a star of mass $m$, and $n_{Lyc} = \Phi_0 \int \tilde{n}_{Lyc}(m) m^{-k}$ is the total production rate of Lyman continuum photons when $r(t) = 1 \ yr^{-1}$. The values of $\tilde{n}_{Lyc}(m)$ are taken from Güsten & Mezger (1983). For the other parameters of the thermal bremsstrahlung we will assume standard numbers:
1) $T_e = 7000 K$, a typical temperature for the electron gas in a HII region (e.g. Osterbrok, 1989)
2) $f_{ion} = 0.8$ (Mezger 1978).

We can therefore parameterize the total thermal radio emission of a galaxy as follows:

$$P_{th}(\nu) = r_0 n_{Lyc}(k) q_{th}(f_{ion}, T_e, \nu) \; , \quad (33)$$

where $q_{th}$ is a known nonlinear function of its arguments.

### 2.3. The FIR/radio ratio

From sections 2.1 and 2.2, the FIR luminosity $L_{\text{FIR}}$ and the radio continuum power $P_\nu$, and consequently the logarithm of their ratio $q_\nu = \log(L_{\text{FIR}}/P_\nu)$, can finally be parameterized as follows:

$$L_{\text{FIR}} = r_0 \, P_{HII} \, \zeta_{HII} \, L_i(k) + r_8 \, P_{UV}(\tau_{UV}) \, \zeta_{diff} \, L_{ii}(k)$$
$$+ r_9 \, P_B(\tau_B) \, \zeta_{diff} \, L_{iii}(k) \quad (34)$$

and

$$P_\nu = P_{th}(\nu) + P_{syn}(\nu)$$
$$= r_0 q_{th}(f_{ion}, T_e, \nu) n_{Lyc}(k) + r_8 N_{SN}(k) q_{SN} \quad (35)$$
$$\Pi_\nu(x, \mu, (U_{\text{rad}}/U_{\text{B}})_i, B_i, D_0, z_h, \delta, \rho, L)$$

In total there are 20 parameters involved in $q_\nu$:
1) Parameters concerning both emissions:
  - the star formation history: $r_0/r_8$
  - the IMF: $k$
2) Parameters concerning the FIR emission:
  - dust absorption: $\tau_B, \tau_{UV}, P_{HII}$
  - dust emission: $\zeta_{diff}, \zeta_{HII}$
  - star formation history: $r_9/r_8$
3) Parameters concerning the radio emission:
  - thermal bremsstrahlung: $f_{ion}, T_e$
  - electron production: $x, q_{SN}$
  - energy losses: $(U_{\text{rad}}/U_{\text{B}})_i, B_i$
  - electron diffusion: $D_0, \mu, z_h$
  - geometry of halo: $L, \rho, \delta$

For some of these parameters, assumptions concerning their values have been made: In section 2.1 we have assumed that $P_{HII} = 0.6$, $\zeta_{HII} = 0.5$ and $\zeta_{diff} = 0.3$, $\tau_B = 0.47 \times \tau_{UV}$, and in section 2.2 we have assumed that $x = 2.2$ (or 2.1), $q_{SN} = 1.6 \ 10^{48}$ eV$^{-1}$, $T_e = 7000$ K, and $f_{ion} = 0.8$. These parameters should be more or less the same for galaxies, and they affect the FIR/radio ratio only weakly. For example, the parameters $P_{HII}$, $T_e$, and $f_{ion}$ depend only on the local conditions for HII regions, and affect only the thermal radio emission and the 'warm' FIR component which are minor components in the radio continuum and FIR emissions. We will therefore in the following take their values as fixed and neglect the influence of their variations on the FIR/radio correlation.

The absolute values of the star formation rate, i.e. $r_0, r_8$ and $r_9$, appear linearly in both expressions of $L_{\text{FIR}}$ and $P_\nu$. Thus $q_\nu$ depends only on the star formation *history*, i.e. on the ratios $r_0/r_8$ and $r_9/r_8$. We will furthermore assume an average value $r_0/r_8 = 1$, corresponding to galaxies with a constant SFR during the past $10^8$ years. We therefore exclude from this analysis starburst galaxies that have had a strongly increased SFR during the last $\approx 10^7$ years.

With these assumptions, there are 11 parameters left for a more detailed discussion. They are: $k$, $\tau_{UV}$, $r_9/r_8$, $(U_{\text{rad}}/U_{\text{B}})_i$, $B_i$, $D_0$, $\mu$, $z_h$, $L$, $\rho$, and $\delta$. If either one of the asymptotic cases, i.e. the 'optically-thick' or 'optically-thin' case discussed in the introduction (Sect. 1), applies for most of the late-type galaxies which follow the FIR/radio correlation, the number of the parameters involved can be further reduced:

(1) *Optically-thick case*: In this asymptotic case $\tau_{UV} \gg 1$ and $|s_h| \gg 1$, i.e. galaxies are optically thick for stellar radiation and synchrotron electrons lose most of their energy before escaping from a galactic halo. Consequently, $P_{UV} \simeq P_B \simeq 1$ in Eq.(34), and the second term on the R.H.S. of Eq.(35), i.e. the synchrotron radiation power $P_{syn}$, is approximated by expression (30). The FIR/radio ratio is therefore a function of only 4 parameters, namely the slope of the IMF k, the star formation history parameter $r_8/r_9$, the strength of the magnetic field B, and the ratio $(U_{\text{rad}}/U_{\text{B}})_i$. If we further ignore the contribution of old stars to the heating of dust (the third term on the R.H.S of Eq.(34)), then the model reduces to the 'calorimeter theory' given by (Völk 1989), where the FIR/radio ratio is inversely proportional to $(1 + U_{\text{rad}}/U_{\text{B}})_i$, and depends only very weakly on the magnetic field strength B and the parameter $k$.

Then, $P_{UV}$ and $P_B \propto \tau_{UV}^{0.9}$ (c.f. Eq.(7)), and $P_{syn}$ is approximated by Eq.(29). The FIR/radio ratio is then a function of 9 parameters: the UV optical depth $\tau_{UV}$, the magnetic field strength B and parameter $\delta$, the four parameters describing the electron propagation: $D_0$, $z_h$, $\mu$, L, and the parameters $k$ and $r_8/r_9$.

## 3. Observational constraints on the parameters

### 3.1. Parameters concerning both emissions

1) For the relevant mass range $1 < m/M_\odot \leq 100$ we adopt a slope of the IMF of $k = 2.5$, consistent with the IMF determined for the solar neighbourhood by Miller & Scalo (1979). Scalo (1986) compared the IMF's of different galaxies and concluded that there is no observational evidence for the slope of the IMF to change between galaxies by more than $\pm 0.5$.

### 3.2. Parameters concerning the FIR emission

2) Assuming a constant gas-to-dust ratio with the same value as observed in our Galaxy, we are able to give an estimate for the average dust opacity $< \tau_{UV} >$ and its dispersion for our galaxy sample. Details are described in Appendix B. The result is $< \tau_{UV} > = 1.1$ with a logarithmic dispersion of $\delta_{\log \tau} = 0.27$.

3) Following the argument of Xu et al. (1994) we assume that the average value of $r_9/r_8 = 1$. Using the formulae given in Xu et al. (1994), we find a dispersion of the logarithm of this ratio of 0.45.

### 3.3. Parameters concerning the radio emission

4) The average values and dispersions of the magnetic field and of the energy density of the radiation field can be estimated for the galaxy sample. The radiation field of a galaxy can be calculated from its bolometric luminosity $L_{bol}$ and its radius $R$. For the magnetic field, only a rough estimate is possible from the total nonthermal radio emission through the minimum energy assumption. The detailed procedure of the calculation of the two quantities is described in Appendix C. For our galaxy sample, the average resulting energy density of the radiation field is $\langle U_{rad} \rangle = 1.4 \mathrm{eV/cm}^3$ and the average value of the magnetic field is $\langle B_{min} \rangle = 7.5 \mu G$ with a dispersion in the logarithm of $\delta_{\log B} = 0.12$.

In Fig. 2 the energy densities of the magnetic field and the radiation field are plotted. Clearly a correlation can be seen. A linear regression analysis yields:

$$U_B \propto U_{rad}^{1.28 \pm 0.12} \quad (36)$$

with a correlation coefficient of $r = 0.74$. The average ratio of the two quantities is

$$\chi = \left\langle \frac{U_{rad}}{U_B} \right\rangle = 1.1. \quad (37)$$

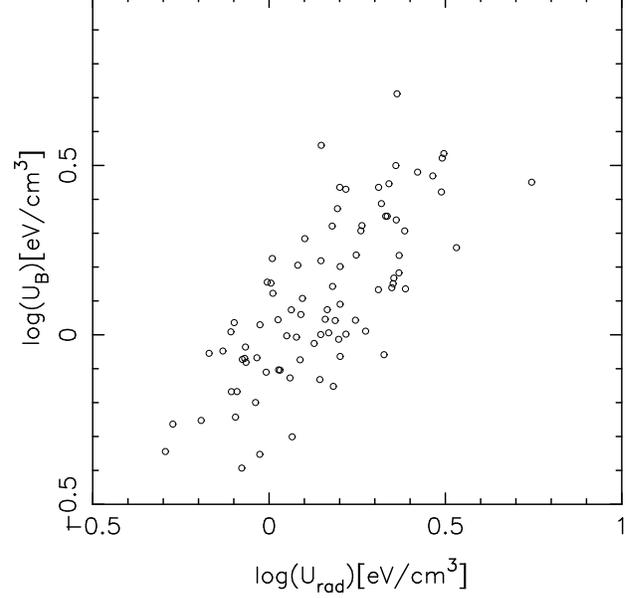

**Fig. 2.** Energy densities of the radiation field and energy densities of the magnetic field, for the galaxy sample

The dispersion in the logarithm of this ratio is $\delta_{\log \chi} = 0.16$. Thus we conclude that the two quantities are reasonably well correlated with a roughly linear relation and that their values are nearly equal. The slight nonlinearity in the relation will not be studied further because it would be an overinterpretation of the simple approach used to calculate $U_{rad}$ (see Appendix C for details).

5) The parameters concerning the electron diffusion ($D_0$, $\mu$, $z_h$) have been estimated from observations for our Galaxy. We will adopt these numbers: $D(1\mathrm{GeV}) = 10^{29} \mathrm{cm}^2/\mathrm{s}$ and $z_h = 10$ kpc (Berezinsky et al. 1990), and the energy dependence of the diffusion coefficient $\mu = 0.5$ (e.g. Berezinsky et al. 1990). This is consistent with a value $x = 2.2$ of the source spectral index (see section 2.2).

6) The extension of the inner halo above the galactic disk is assumed to be $L = 1 kpc$. This value lies within the range determined for the extension of the thin radio disk for the edge-on galaxies NGC 819 (L=400 pc, Dahlem, Dettmar & Hummel 1994) and NGC 4631 (L=1.5 kpc, estimated from the maps by Golla & Hummel 1994).

The parameter $\rho$, describing the decrease of the radiation energy density away from the galactic disk is given by the geometrical dilution. With the above values of $L$ and $z_h$, we derive $\rho = 0.4$.

The decrease of the magnetic field in the halo is more difficult to estimate. For our Galaxy, dynamo theory predicts a ratio of the magnetic field strength in the halo, $B_h$, to that in the disk, $B_d$, of $B_h/B_d \approx 0.1$ (Ruzmaikin et al. 1988) In other galaxies this ratio might be higher: For NGC 4631 a ratio of $B_h/B_d \approx 1$ is conceivable (Hummel, Beck & Dahlem 1991). We will consider various values for $\delta$ in this range.

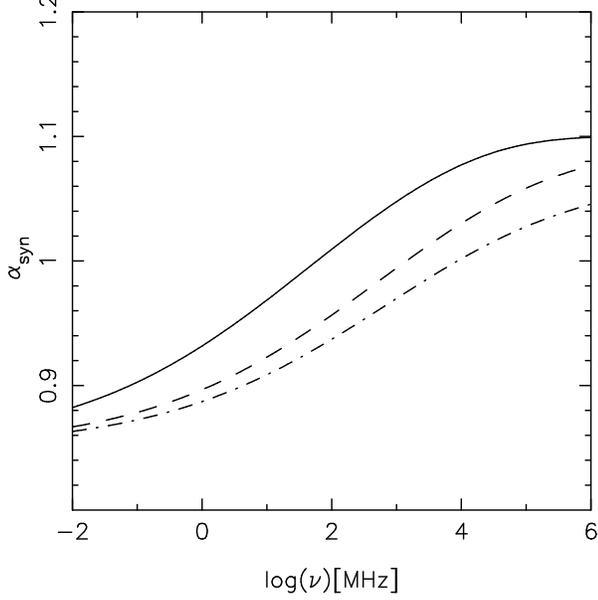

**Fig. 3.** Spectral index $\alpha_{syn}$ of the synchrotron radiation shown for $\delta = 1, \rho = 1$ (full line), $\delta = 0.5, \rho = 0.4$ (dashed line) and $\delta = 0.2, \rho = 0.4$ (dashed-dotted line). $B_i = 7\mu G$, $(U_{rad}/U_B)_i = 1.1$, according to the results of Sec. 2.2 and $z_h = 10$kpc, $D_0 = 10^{29} cm^2/s$, according to the values in our Galaxy, and $L = 1$kpc are adopted.

A further constraint to the parameters concerning the synchrotron emission, is given by the nonthermal spectral index, $\alpha_{syn}$. Klein (1988) separated for 13 galaxies the thermal from the nonthermal radio emission and derived an average nonthermal spectral index of $0.88 \pm 0.06$. Independently, Duric, Bourneuf & Gregory (1988) obtained an essentially identical result.

The prediction for the spectral index within this model is given in Fig. 3 for the parameters listed above. In addition to $\delta = 0.2, \rho = 0.4$ and $\delta = 0.5, \rho = 0.4$ we have plotted for comparison the case of a homogeneous halo and disk ($\delta = \rho = 1$). The predicted spectral index of the synchrotron radiation between 100 MHz and 10 GHz lies between $1.0 - 1.1$ in the uniform case, and $0.9 - 1.0$ for an outwardly decreasing magnetic field. With the slightly lower source spectral index ($x = 2.1$) advocated by Berezkho et al. (1993), these values decrease to $0.95 - 1.05$ in the homogeneous case, and to $0.85 - 0.95$ for a decreasing magnetic field. The decrease of the magnetic field in the halo produces a lower spectral index which is in reasonable agreement with the observations, indicating that the parameter values we have chosen are in the right range.

## 4. Dispersion of the FIR/radio correlation within the model

For our galaxy sample we find a mean of the logarithm of the FIR/radio ratio $q_{1.49GHz} = 15.06$ with a dispersion of $\sigma_{q_{1.49GHz}} = 0.24$. This can be compared to the dispersion derived within the model: We calculated the derivative of $q_{1.49GHz}$ with respect to the logarithms of various parameters, represented by $x_i$ (i=1, 2, ...), using the mean values of the parameters discussed in the previous section as input. This number gives a quantitative estimate how sensitive the FIR/radio ratio depends on the respective parameter. For some of the parameters we could derive the mean value and its dispersion, $\delta_{\log(x_i)}$, and thus we can give an estimate for their contribution, $\sigma_{x_i}$, to the scatter in the logarithm of the FIR/radio ratio $q_{1.49GHz}$ by:

$$\sigma_{x_i} = \frac{\partial q_\nu}{\partial \log(x_i)} \delta_{\log(x_i)}, \tag{38}$$

where the derivative is taken at the radio frequency of 1.49 GHz. Table 1 summarizes the results (for $\delta = 0.2, \rho = 0.4$). The parameter $k$ is not included because its contribution to the dispersion of the correlation is not significant ($\sigma_k < 0.1$ for $\delta_k < 0.5$). The parameters $\delta, \rho, L$ and $\mu$ are also dropped because their influence on the synchrotron emission is only second-order and therefore less important compared to $(U_{rad}/U_B)_i$, $B_i$, $z_h$ and $D_0$.

**Table 1.** Contributions to the dispersion of the FIR/radio correlation by individual parameters

| $x_i$ | $<x_i>$ | $\partial \frac{q_{1.49GHz}}{\partial \log(x_i)}$ | $\delta_{\log(x_i)}$ | $\sigma_{x_i}$ |
|---|---|---|---|---|
| $\tau_{UV}$ | 1.1 | 0.4 | 0.27 | 0.11 |
| $B_i$ | 7.5$\mu$G | -0.9 | 0.12 | -0.11 |
| $(U_{rad}/U_B)_i$ | 1.1 | 0.3 | 0.16 | 0.05 |
| $r_9/r_8$ | 1 | 0.3 | 0.45 | 0.14 |
| $D_0$ | $10^{29} cm^2/s$ | 0.5 | | |
| $z_h$ | 10 kpc | -0.4 | | |

For the first four parameters in Table 1, namely the UV optical depth $\tau_{UV}$, the strength of the magnetic field in the inner thin disk $B_i$, the ratio between the energy densities of the radiation field and that of the magnetic field $(U_{rad}/U_B)_i$, and the star formation history parameter $r_9/r_8$, quantitative estimates of their contributions to the scatter of the FIR/radio ratio are presented. The absolute values are all smaller than the dispersion of $q_{1.49GHz}$. The contribution from the parameter $(U_{rad}/U_B)_i$ is negligibly small, due to the reasonably tight correlation between $U_{rad}$ and $U_B$ (see section 3.3) and a relatively weak dependence of $q_{1.49GHz}$ on this parameter (the logarithmic derivative equals to 0.3). The contributions of $D_0$ and $z_h$ are not given, because we do not have the data to evaluate their dispersions although, according to our calculation, the FIR/radio ratio depends on these two parameters rather sensitively (the logarithmic derivatives are 0.5 and 0.4, respectively).

In principle the variance, i.e. the square of the dispersion of $q_{1.49GHz}$ is given by:

$$\sigma^2_{q_{1.49GHz}} = \sum_i \sigma^2_{x_i} + 2 \sum_{i,j}^{i<j} \sigma_{x_i} \sigma_{x_j} r_{i,j}. \tag{39}$$

The second term on the R.H.S. takes into account correlations between the parameters $x_i$ and $x_j$, where $r_{i,j}$ is the linear correlation coefficient between $x_i$ and $x_j$. In Table 2 we list the contributions of the cross correlations between the three parameters $\tau_{UV}$, $B$, and $r_9/r_8$, for which we find a significant $\sigma$ in Table 1. Significant anticorrelations between $\tau_{UV}$ and $r_9/r_8$, and between $B$ and $r_9/r_8$ are found, which result in significant contributions (with opposite signs) to the variance of $q_{1.49GHz}$ (Table 2). The anticorrelation between the optical depth and $r_9/r_8$ may be understood in terms of the well established correlation between the gas column density and the present day star formation rate (e.g. Buat 1992), given that our $\tau_{UV}$ is estimated from the HI column density. The physical interpretation of the second correlation, i.e. the $B$ vs. $r_9/r_8$ anticorrelation, may be drawn from the dynamo theory for the galactic magnetic field (Ko & Parker 1990), which predicts an enhancement in B for a more active present day star formation activity (i.e. a lower $r_9/r_8$). No significant correlation is found between $\tau_{UV}$ and $B$ although both are correlated with $r_9/r_8$.

**Table 2.** Contributions to the variance of the FIR/radio correlation by three cross correlations

|  | $r_{i,j}$ | $2\sigma_{x_i}\sigma_{x_j}r_{i,j}$ |
| --- | --- | --- |
| $\log \tau_{UV}$ v.s. $\log B$ | -0.015 | 0.0004 |
| $\log \tau_{UV}$ v.s. $\log(r_9/r_8)$ | -0.50 | -0.015 |
| $\log B$ v.s. $\log(r_9/r_8)$ | -0.62 | 0.021 |

From the values in Table 1 and Table 2 we can predict a lower limit of $\sigma^2_{q_{1.49GHz}} > 0.052$. This is consistent with the observed value $\sigma^2_{q_{1.49GHz}} = (0.24)^2 = 0.058$, leaving some room for the unknown contribution from the parameters $D_0$ and $z_h$. The absolute value of $q_{1.49GHz}$ predicted by our model is $q_{1.49GHz} = 14.6 \pm 0.3$, slightly lower than the observed value $q_{1.49GHz} = 15.06 \pm 0.24$. Several uncertainties in our model may be responsible for this discrepancy: (1) A lower-mass cut-off of $8M_\odot$, instead of the $5M_\odot$ used in our calculation for the SN rate $\nu_{SN}$, will raise the prediction for $q_{1.49GHz}$ to $14.9 \pm 0.3$; (2) the error in the derivation of $q_{SN}$. It should be pointed out, however, that our results on the dispersion of $q_{1.49GHz}$ are not at all affected by this discrepancy.

## 5. Optically-thick or optically-thin?

With the parameters as they are determined using our galaxy sample, we now can discuss the question whether most late-type galaxies, as represented by our sample, are 'optically-thick' or 'optically-thin' for their primary photons and CR electrons.

The mean optical UV depth $\tau_{UV} = 1.1$ with the logarithmic dispersion of 0.27 (a factor of 1.86 in linear scale) indicates that most of galaxies are *optically thick*, although not extremely so, for the UV radiation which is most responsible for the heating of dust (Xu 1990; Xu & Buat 1995). Our result is based on the dust-to-gas ratio adopted from the value in the solar neighborhood. The validity of this assumption has recently been confirmed by Xu & Buat (1995) and Buat & Xu (1995) who studied the extinction in late-type galaxies using a radiative transfer model which allows for the effect of dust scattering and which takes input from the UV, optical and FIR observations of a sample of nearby late-type galaxies. Xu & Buat (1995) found a mean optical depth for the blue radiation (4400Å) $\tau_B = 0.6 \pm 0.04$, indicating a $\tau_{2000\text{Å}} = 1.08 \pm 0.07$, consistent with our result.

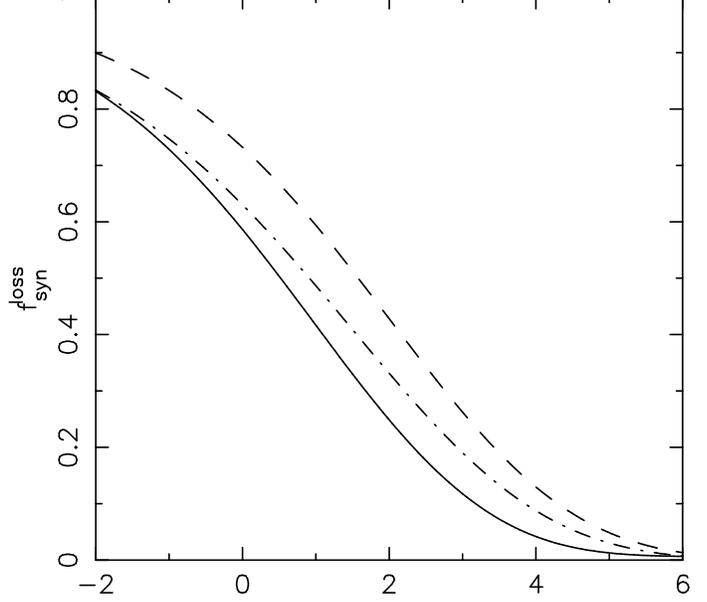

**Fig. 4.** The fraction of synchrotron emission lost due to electron escape, $f^{loss}_{syn}$, as a function of frequency, for the same parameters as in Fig.3

Within our theory, we can also quantify the importance of escape of CR electrons: In Fig. 4 the fraction of synchrotron emission lost due to electron escape $f^{loss}_{syn}$ is plotted.

It is defined as:

$$f^{loss}_{syn} = 1 - \frac{P_{syn}}{P^{max}_{syn}} \qquad (40)$$

with $P^{max}_{syn}$ being the maximum synchrotron emission for given parameters $L, \delta$ and $\rho$, assuming an infinite extent of the halo. At 1 GHz the fraction of synchrotron emission lost is between $10 - 25\%$. The dominant part of the energetically possible synchrotron emission is thus observed. This result is confirmed by the value for $|s_h|$ that we derive in the model: At 1 GHz we obtain (in the spatially homogenous case) a value of $|s_h| = 0.69$ which means that the diffusion time scale $\tau_{diff}$ is 5.5 times longer than the energy loss time scale $\tau_{loss}$ (see Eq.(23)).

Therefore we conclude that the 'optically-thick' scenario for the FIR/radio correlation describes the situation for late-type galaxies better than the 'optically-thin' scenario. On the other hand the extreme conditions assumed in the 'calorimeter theory', namely $\tau_{UV} >> 1$ and $f^{loss}_{syn} << 1$, are not strictly fulfilled, as can also be seen from the logarithmic derivatives given in Table 1. In the extreme case, one should have nearly zero logarithmic derivatives with respect to $\tau_{UV}$, $D_0$, and $z_h$, and a very small derivative with respect to $B$ (Völk 1989). Yet we find that the absolute values of these derivatives are all significantly larger than zero (Table 1). Nevertheless, these values are indeed significantly smaller than they would be in the 'optically-thin' case, namely 0.9 for $\tau_{UV}$ (Eq(7)), 1 for $D_0$ and 2 for $z_h$ (Eq(29)), and 1.85 for $B$. In the optically thin case, the stronger dependence of $q_\nu$ on the various parameters would result in a higher dispersion: Taking into account only the dispersion for $B$, $\tau_{UV}$ and $r_9/r_8$ (thus neglecting $D_0$ and $z_h$) we would already predict $\sigma^2_{1.49GHz} = 0.13$, exceeding by far the observed dispersion of the FIR/radio correlation. Therefore, as shown in the previous section, the tight FIR/radio correlation can actually be explained by these weak dependencies, as has been argued in the 'optically-thick' scenario, without invoking any unknown relations among the parameters unless the contributions of $D_0$ and $z_h$ to the dispersion are very significant. Here, the tight correlation between $U_{\rm rad}$ and $U_{\rm B}$ is very important for the low predicted value of $\sigma^2_{1.49GHz}$, reducing the contribution of $(U_{\rm rad}/U_{\rm B})_i$ to a very small value. Therefore, the proportionality between $U_{\rm rad}$ and $U_{\rm B}$ can be considered as a necessary condition for the existence of the tight FIR/radio correlation. Furthermore, we do not find any significant correlation between $\tau_{UV}$ and $B$, indicating the absence of a correlation between the trapping rate of the primary photons and that of CR electrons as required by the 'optically-thin' scenario.

## 6. Discussion

### 6.1. Model simplifications

In the model describing the radio and the FIR emission, several simplifications were introduced, the major ones being the use of a 1-dimensional diffusion model to describe the electron propagation and the assumption of an infinite slab geometry for the galactic disk in which stars and dust are uniformly mixed. In the following we will discuss these assumptions.

By using a slab geometry for the model of the dust heating we assume that the scale heights for dust and stars are the same. This is most likely a good approximation for massive stars responsible for the UV-radiation ($m \gtrsim 5 M_\odot$), whereas the scale height of lower mass stars is larger than that of the dust. Therefore the FIR rises less with the dust opacity because less radiation from the old population is converted into FIR emission. This smaller influence of the old stellar population on the FIR/radio ratio results in a correspondingly smaller effect of the long time star formation history on the FIR/radio ratio. However, since the UV radiation of massive stars is the dominant heating source of dust per stellar mass unit (Xu, 1990), we expect this effect to be noticeable only for galaxies with very low SFR during the past $10^8$ yrs.

The uniform mixing of stars and dust probably leads to a decrease of the overall FIR production in a disk galaxy rather than to an increase. The reason is that the dominant massive stars generally form and evolve in regions of increased dust density. This should on average *increase* the overall galactic optical depth for their UV radiation, and thus make galaxies overall more "optically-thick" than does uniform mixing (section 5).

A 1-dimensional model describes the electron propagation in a galaxy correctly if the distribution of the sources of CR's is uniform within the galactic disk. In this case radial diffusion in the disk plays no role (except for regions close to the radial disk boundaries) and the electron phase space density, f, is only a function of z and E. As long as diffusion is a linear process (no effect of the cosmic rays on their own diffusion), this appears a reasonable approximation for the *integrated* radio emission of a galaxy. If we were to investigate *local* correlations between the FIR and the radio emissions, clearly a more realistic geometry for the distribution of stars and SNe would have to be taken into account, and a three-dimensional diffusion model would be necessary.

The synchrotron emission was calculated in a diffusion model, not taking into account other modes of CR propagation. However, convection can also play a role in the transport of CRs, especially at large distances from the disk: For example, the observations of the halo radio emission of NGC 4631 can be explained by convection (Hummel & Dettmar 1990). Convection will be caused by a galactic wind which should exist even for a galaxy like ours (Breitschwerdt, McKenzie & Völk 1991, 1993). It should certainly dominate diffusion in starburst galaxies like M82.

If radiative energy losses dominate the distribution function of electrons, the spatial transport mechanism (diffusion or convection) affects the integrated spectral index only if the magnetic field is spatially inhomogeneous. Otherwise (spatially homogeneous $B$, dominant energy losses), the transport mechanism is irrelevant for the integrated synchrotron emission. In the general case, convection will tend to produce a flatter spectrum and a slightly weaker dependence of the synchrotron emission on the magnetic field than does purely diffusive transport. This can be seen from the following argument: convection is an energy independent transport mechanism. Therefore, if the CR propagation were only by convection, in the asymptotic case of very low frequencies the spectral index $\alpha$ would be the same as in Eq.(29) for $\mu = 0$

$$\alpha_{\rm conv} = \left(\frac{x-1}{2}\right) = 0.6 \text{ (for } x = 2.2\text{)}. \tag{41}$$

The logarithmic derivative $\beta$ of the synchrotron emission with respect to B for $\mu = 0$ is for very low frequencies,

$$\beta_{\rm conv} = \frac{x+1}{2} = 1.6. \tag{42}$$

Thus if convection plays a role, the spectral index will be, also at finite frequencies, in even better agreement with the observations, because of the flatter asymptotic spectrum of $P_{syn}(\nu)$. The conclusions about the influence of the magnetic field will, on the other hand, not be changed significantly.

### 6.2. Correlation between $U_{\rm rad}$ and $U_{\rm B}$

An important result of this paper is the tentative discovery of a near proportionality between the energy density of the radiation field and that of the magnetic field. In turn it could be derived from the empirical existence of the FIR/radio correlation as a necessary condition for it. This additional correlation is extremely interesting by itself. A qualitative physical explanation for this correlation could be in terms of turbulent dynamo activity in galaxies: The emission from massive stars dominates the interstellar radiation field. Through stellar winds and SN explosions these stars produce at the same time interstellar turbulence whose hydrodynamic energy density, $U_{turb}$, is transported with hydrodynamic velocities that are $10^2$–$10^3$ times smaller than the velocity of light. This can compensate for the fact that for a quasi-stationary galaxy the time-averaged hydrodynamic energy flux into interstellar space is smaller by a similar factor than the photon luminosity from the stars. Thus the hydrodynamic energy density $U_{turb}$ may be comparable to the energy density $U_{\rm rad}$ of the interstellar radiation field, as it is observed in our Galaxy. In turn, the equipartition of turbulent and magnetic pressures limits the dynamo effect that increases the magnetic field and therefore can lead to a near equipartition between $U_{turb}$, $U_{\rm rad}$ and $U_{\rm B}$.

A detailed model based on these ideas will be given in a future paper.

## 7. Conclusions

We conclude that the FIR/radio correlation is due to a few basic properties of late type galaxies like a sufficient optical thickness of their gas/dust disks, a sufficiently large size of their magnetized halos, and a correlation between the energy densities of the magnetic and the radiation field, most likely due to a near equipartition between random kinetic energy density, and energy density of the interstellar photon field. Probably the disruptive tendency of energy production as a consequence of star formation is just in dynamical equilibrium with the tendency of the gaseous disk and halo to contract gravitationally, and to produce stars at a higher rate. Similarly, the energy loss of the interstellar medium due to radiative cooling and due to the escape of the CR nucleon component (energetically by far dominating the radiatively cooling CR electron component) together with a galactic wind, just appears large enough to keep the various energy densities $U_{turb}$, $U_{\rm rad}$, and $U_{\rm B}$ at a comparable level.

In this paper we have only been partly able to explain these deep connections. Rather we have theoretically tied together their empirical consequences, like the values of optical depths, etc.. One of these consequences is the FIR/radio correlation. It theoretically requires the existence of an $U_{\rm rad}/U_{\rm B}$ correlation which was also found empirically if we accept the minimum energy assumption as an approximation to calculate galactic magnetic fields. To this extent the $U_{\rm rad}/U_{\rm B}$ correlation is the major discovery of this paper. In principle this correlation allows the determination of galactic magnetic field strengths from observations of the FIR/radio ratio, and of the integrated photon flux from a galaxy of known source geometry. In practice many other parameters enter this calculation. Nevertheless it is a rare case in astrophysics that observations plus theory directly imply new astrophysical relations which were not known before but are at least vaguely plausible as we have speculated above.

Implicit in all the estimates we have presented is the dominant role of SN precursors and their ultimate explosions for the dynamics of the interstellar medium in most normal late-type galaxies. However there must be a limit to such a behavior. If the current ($\sim 10^8$ yrs) star formation rate becomes too small compared to the average rate over times of the order of $10^9$ yrs, then old stars will dominate the FIR radiation without contributing to the radio continuum emission as argued by Condon et al. (1991) and Xu et al. (1994). Thus there is also no physical reason any more for a correlation between $U_{\rm rad}$ and $U_{\rm B}$. Such galaxies will probably not differ much in their structure from the more active ones. But their radiation fields in different wavelength ranges will not be correlated, because the energetically dominant stellar radiation from old stars does not strongly couple to the dynamics of the interstellar medium. From this point of view they are less interesting. As a consequence the next question regards the opposite case of starburst galaxies which are characterized by variations on time scales $< 10^8$yrs. They also appear to follow the FIR/radio correlation although it is not immediately clear why this should be the case. This question will be discussed in a companion paper.

*Acknowledgements.* We would like to thank V. Dogiel for stimulating discussions about CR propagation, E. Hummel for his helpful advice on analysing the $U_{\rm rad} - U_{\rm B}$ correlation, and E. Wunderlich for selecting the galaxy sample. We would also like to thank the referee for very useful and detailed comments. UL gratefully acknowledges the receipt of a stipend of the Max-Planck-Gesellschaft. HJV and CX acknowledge that part of their work in this paper has been done in the framework of the Sonderforschungsbereich 328 (Entwicklung von Galaxien) of the Deutsche Forschungsgemeinschaft.

## Appendix A: Solution of the diffusion equation

In the asymptotic cases $|s_h| \gg 1$ and $|s_h| \ll 1$ Eq. (18) is an exact solution to the diffusion equation Eq.(11). In the range between these asymptotes, the function $f(E,z)$ given by Eq. (18) does not fulfill Eq. (11) exactly, but is a good approximation. Its accuracy was estimated in the following way: Eq. (18) satisfies exactly the equation:

$$D(E)\frac{\partial^2 f(E,z)}{\partial z^2} + \frac{\partial}{\partial E}\left(b(E)f(E,z)\right) = \\ -\left(\frac{E}{m_e c^2}\right)^{-x} \nu_{\rm SN} q_{\rm SN} \left[\delta(z) + \triangle(E,z)\right] \ ,$$

where the residue $\triangle(E,z)$ equals:

$$\triangle(E,z) = 2\frac{s_h}{z_h} {}_1F_1(p, \frac{1}{2}, s) \\ \left( s_h \frac{{}_1F_1(p+\frac{1}{2}, \frac{3}{2}, s_h)}{{}_1F_1(p, \frac{1}{2}, s_h)^2} \frac{\partial {}_1F_1(p, \frac{1}{2}, s_h)}{\partial s_h} - \right. \\ \left. \frac{1}{2}\frac{{}_1F_1(p+\frac{1}{2}, \frac{3}{2}, s_h)}{{}_1F_1(p, \frac{1}{2}, s_h)} - \frac{s_h}{{}_1F_1(p, \frac{1}{2}, s_h)} \frac{\partial {}_1F_1(p+\frac{1}{2}, \frac{3}{2}, s_h)}{\partial s_h} \right)$$

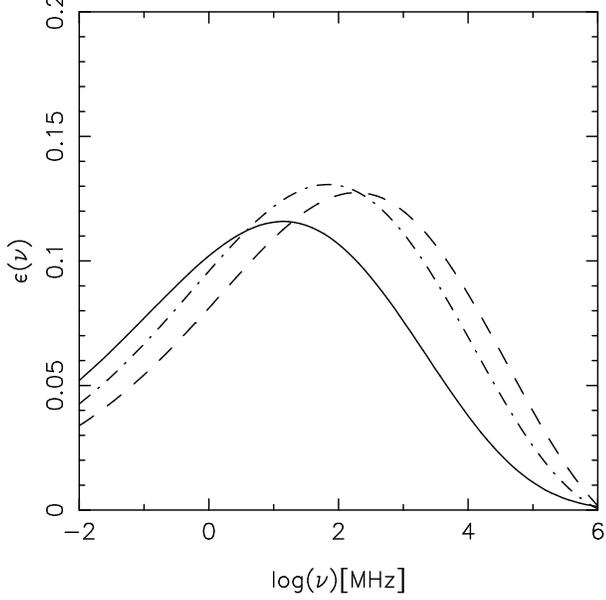

**Fig. A1.** The integral over the residue $\int_{-z_h}^{z_h} \triangle(E,z)\mathrm{d}z$. The parameters describing the halo geometry are the same as in Fig. 3, i.e. $L/z_h = 0.1$, $\delta = \rho = 1$ (full line), $\delta = 0.5$, $\rho = 0.4$ (dashed line) and $\delta = 0.2$, $\rho = 0.4$ (dashed-dotted line).

For the case of $\delta < 1$, or $\rho < 1$, an analogous (but more complicated) expression for $\triangle(E,z)$ can be derived. The expression:

$$\varepsilon(\nu) = \frac{\int_{-z_h}^{z_h} \triangle(E(\nu),z)\mathrm{d}z}{\int_{-z_h}^{z_h} \delta(z)\mathrm{d}z} = \int_{-z_h}^{z_h} \triangle(E(\nu),z)\mathrm{d}z$$

gives an estimate for the error of the approximate solution. $\varepsilon(\nu)$ is shown in Fig. (A1), both for the spatially homogeneous as well as for the inhomogeneous case. It can be seen that the accuracy in the homogeneous case is always better than 12 %, and in the inhomogeneous case it is better than 14 %. The error of $\alpha_{syn}$ and of the derivatives of $q_{1.49GHz}$ listed in Table 1 is much lower: it is for all quantities less than 2 %.

## Appendix B: Dust opacity

With the data available for the galaxy sample, we want to calculate the dust opacity and its dispersion. The sample of 114 normal spiral galaxies that we use in this paper is the same as used and described in Xu et al. (1994). It is selected from the catalogue of Aaronson et al. (1982) of nearby late-type galaxies with known H-band (1.6$\mu$m) magnitude, for which also FIR and radio fluxes are available.

We are interested in the diffuse dust, associated with HI, which in our Galaxy has been shown to emit a large fraction of the total FIR emission (Boulanger & Pérault 1988, Cox & Mezger 1988, Pérault et al. 1988, Bloemen, Deul & Thaddeus 1990). The FIR emission itself is a poor indicator of the dust mass because it depends very sensitively on the, not very well known, dust temperature distribution. More importantly, it does not allow a distinction regarding the origin of the dust emission (i.e diffuse or from HII region). Therefore we take a different approach: We assume that the dust-to-gas ratio, which for our Galaxy is observed to be quite constant (e.g. Savage & Mathis 1979), is the same in other galaxies. For some nearby galaxies this assumption has been shown to be valid: Rice et al. (1990) could fit the infrared emission and the optical and UV spectrum of M33, using the Galactic value of the dust-to-gas ratio. For M31, Walterbos & Kennicutt (1988) showed that in M31 the value is also the same as in our Galaxy. In the very metal poor Large Magellanic Cloud, the ratio is higher by a factor of 4 (Koornneef 1982) however, and in the Small Magellanic Cloud even higher by a factor of 8-12 (Bouchet et al., 1985). Since the galaxy sample used here consists mostly of spiral galaxies, we believe that a constant gas-to-dust ratio is a reasonable assumption in this case. We take N(HI)/E(B − V) = 5 · $10^{21}$atoms/$cm^2$mag (Savage & Mathis 1979) with N(HI) being the column density of HI, and E(B-V) the color excess. This gives an average absorption cross section per H-atom:

$$\sigma_V = \frac{R}{1.086}\frac{E(B-V)}{N(HI)} = 5.71 \cdot 10^{-22}\mathrm{cm}^2/\mathrm{atom} \quad (B1)$$

with $R = 3.1$ (Savage & Mathis 1979). The visual dust opacity $\tau_V$ in a galaxy can therefore be estimated with the use of $\sigma_V$, the HI mass F(HI), and the diameter of the galaxy D:

$$\tau_V = \sigma_V \frac{F(HI)}{m_p} \frac{1}{\pi(D/2)^2}. \quad (B2)$$

$m_p$ is the proton mass. The opacity in the UV is higher: The average extinction curve of dust (Savage & Mathis 1979) shows that the extinction at 912 Å is a factor of 5.72 higher than in the visual wavelength range. For the galactic diameter we can use the optical isophotal diameter $D_{25}$ instead of the HI diameter with a small correction factor: Fouqué (1983) found that the effective HI diameter $D_H$, i.e. the diameter within which half of the HI mass is situated, and $D_{25}$ are nearly the same: The ratio $D_H/D_{25} = 1 - 1.5$ for spiral galaxies, depending on morphological type, with a very low dispersion of less than 0.1 in the logarithm. Thus we approximate the FIR disk diameter by $D_{25}$ and apply the morphological type dependent correction factor of Fouqué.

The average opacity determined in this way at 2000 Å is $<\tau_{UV}> = 1.1$ with a dispersion in the logarithm of $\delta_{\log \tau} = 0.27$. There is no tendency noted for the opacity to increase or decrease with luminosity. Thus the galaxies are optically thick for the non-ionizing UV light which is mainly emitted by massive stars. In the optical wavelength range galaxies are optically thin ($<\tau_V> = 0.4$). Therefore preferentially the radiation from massive stars is absorbed.

## Appendix C: Calculation of the energy density of the radiation field and of the magnetic field for the galaxy sample

Assuming that the radiation intensity in the disk is homogeneous and isotropic, the energy density of the radiation field in the galactic disk is given by:

$$U_{\mathrm{rad}} = \frac{2L_{bol}}{c\pi(D_{25}/2)^2} = \frac{(L_{bol}/W)}{(D_{25}/\mathrm{kpc})^2} \cdot 5.56 \cdot 10^{-35} \left[\frac{eV}{cm^3}\right] \quad (C1)$$

with $D_{25}$ being the optical isophotal diameter at 25 mag, and $L_{bol}$ the bolometric luminosity. The photons contributing to the inverse Compton losses have to be below the Klein-Nishina limit which means that they have to follow the condition:

$$\frac{h\nu}{mc^2} \frac{E}{mc^2} \ll 1$$

(e.g. Longair 1992). For electrons in the energy regime considered here (about $1 - 20$ GeV) the upper limit lies in the X-ray range ($h\nu \approx 200$ eV). Data for the luminosities at various wavelength bands are available: the blue and H magnitudes and the FIR emission. With these data, and taking into account the microwave background, the total radiation energy density of a galaxy can be estimated by extrapolation. From the blue magnitude the magnitudes in the visual and in the UV are estimated by the average (U-B) and (B-V) values corresponding to each morphological type (Huchra 1977). Obtaining in this way the luminosities per wavelength at UV, visual, blue wavelength and at 1.6 $\mu$m (H-band), one can estimate the total luminosity between 912 Å and 4 $\mu$m by extrapolation, referring to the calculated galaxy spectra in Mazzei, Xu & De Zotti (1992).

$$L_{912-4000}[W] = L_{UV}[W/\mu m] \cdot 0.2\mu m + L_B[W/\mu m] \cdot \\ 0.1\mu m + L_V[W/\mu m] \cdot 0.1\mu m + L_H[W/\mu m] \cdot 4\mu m. \quad (C2)$$

In addition we take into account the FIR luminosity ($40 - 120\mu$m) and the microwave background with an energy density of $U_{\mathrm{back}} = 0.25$eV/cm$^3$. The wavelength range between $4\mu$m and $40\mu$m is negligible for the total luminosity of a galaxy. The total radiation energy density is the sum of the energy densities at different wavelengths. All these energy densities individually correlate positively that the energy density of the magnetic field, but with very different slopes (between 0.6 and 1.7). As a consequence, the slope in the correlation between $U_{\mathrm{rad}}$ and $U_{\mathrm{B}}$ depends rather sensitively on the weight which each luminosity has in Eq. (B1). The correlation coefficient and the dispersion of the correlation, on the other hand, are not significantly affected by it.

The magnetic field is estimated by the standard minimum energy approximation (e.g. Pacholczyk 1970). It is based on the assumption that the value of the magnetic field is such that the sum of the CR energy and the magnetic field energy is minimal. The magnetic field can in this way be calculated from the synchrotron emission $P_{\mathrm{syn}}(\nu)$ and the volume of the radio emitting region:

$$B_{min} = 4.2 \cdot 10^{-5} \left(\frac{(P_{\mathrm{syn}}(\nu)/\mathrm{W/Hz})(\nu/\mathrm{GHz})^\alpha}{(D/\mathrm{kpc})^2}\right)^{\frac{2}{7}} [\mu G].$$

In this formula we used the standard parameters: The proportionality factor between the energy of the cosmic ray electrons and the total energy of the cosmic rays is $k = 100$, the spectral index of the synchrotron emission $\alpha = 0.7$ and the minimum and maximum frequency of the synchrotron emission $\nu_{min} = 10$ MHz and $\nu_{max} = 10$GHz. The radio emitting volume is taken to be a cylinder of 1 kpc height. Its diameter $D$ is given by the lowest contour line of the galaxy maps given in Condon (1987) and Condon, Yin & Burstein (1987). Only 86 galaxies of the sample could be considered because for the rest no diameter could be deduced.

*Acknowledgements*. We would like to thank V. Dogiel for stimulating discussions about CR propagation, E. Hummel for his helpful advice on analysing the $U_{\mathrm{rad}} - U_{\mathrm{B}}$ correlation, and E. Wunderlich for selecting the galaxy sample. We would also like to thank the referee for very useful and detailed comments. UL gratefully acknowledges the receipt of a stipend of the Max-Planck-Gesellschaft. HJV and CX acknowledge that part of their work in this paper has been done in the framework of the Sonderforschungsbereich 328 (Entwicklung von Galaxien) of the Deutsche Forschungsgemeinschaft.

## References

Aaronson M., Huchra J., Mould J.R., et al., 1982 ApJS , 50, 241
Berezinsky V.S., Bulanov S.V., Ginzburg V.L., Dogiel V.A., Ptuskin V.S., 1990, *Astropysics of Cosmic Rays*, North Holland
Berezkho E.G., Yelshin V.K., Ksenofontov L.T., 1993, Proceedings of the 23rd International Cosmic Ray Conference, Vol. 2, p. 354
Beuermann K., Kanbach G., Berkhuijsen E.M., 1985, A&A 153, 17
Boulanger F., Pérault M., 1988, ApJ 330, 964
Bouchet P., Lequeux J., Maurice E., Prévot L., Prévot-Burnichon M.L., 1985 A&A 149, 330
Bloemen J.B.G.M., Deul E.R., Thaddeus P., 1990, A&A 233, 437
Breitschwerdt D., McKenzie J.F., Völk H.J., 1991, A&A 245, 79
Breitschwerdt D., McKenzie J.F., Völk H.J., 1993, A&A 269, 54
Buat V., 1992, A&A 264, 44
Buat V., Xu C., 1995, A&A , submitted
Cesarsky C., Montmerle T., 1983, Space Sci. Rev. 36, 173
Chi X., Wolfendale A.W., 1990, MNRAS 245, 101
Condon J.J., 1987, ApJS 65, 485
Condon J.J., Yin Q.F., Burstein D., 1987, ApJS 65, 543
Condon J.J., Anderson M.L., Helou G., 1991 ApJ , 376, 95
Condon J.J., 1992, A&AR 30, 575
Cox P., Mezger P.G., 1988, in *From Comets to Cosmology*, ed. E. Lawrence, Third International IRAS Conference, p.97
Dahlem M., Dettmar R.-J., Hummel E., 1994 A&A 290, 384
Désert F.X., Boulanger F., Puget J.L., 1990, A&A 237, 215
Duric N., 1986, ApJ 304, 111
Duric N., Bourneuf E., Gregory P.C., 1988, ApJ 96, 81
Fouqué P., 1983, A&A 122, 273
Ginzburg V.L., Syrovatskii S.I., 1964, *The Origin of Cosmic Rays*, Pergamon Press, Oxford
Golla G., Hummel E., 1994, A&A 284, 777
Güsten R., Mezger P.G., 1983, Vistas in Astronomy 26, 159
Helou G., 1991, in *The Interpretation of Modern Synthesis Observations of Spiral Galaxies*, eds. N. Duric and P. Crane, PASP Conf. Ser. (in press)
Helou G., Bicay M.D., 1993, ApJ 415, 93
Huchra J.P., 1977, ApJS 35, 171
Hummel E., Dettmar R.-J., 1990, A&A 236, 33
Hummel E., Beck R., Dahlem M., 1991, A&A 248, 23
Klein U., 1988, *Radio Emission and Star Formation in Spiral and Dwarf Irregular Galaxies*, Habilitationsschrift, Bonn
Ko C.M., Parker E.N., 1989, ApJ , 341, 828

Koornneef J., 1982, A&A 107, 247
Lisenfeld U., 1993, PhD thesis (in German), University of Heidelberg
Longair M.S., 1992, *High Energy Astrophysics*, Vol. 1, Cambridge University Press
Markiewicz W.J., Drury O'C., Völk H.J., 1990, A&A 236, 487
Mathews J., Walker R.L., 1970, *Mathematical Methods of Physics*, 2nd ed., Benjamin Cumming Publishing Company
Mathis J.S., Mezger P.G., Panagia N., 1983 A&A 128, 212
Mazzei P., Xu C., De Zotti G., 1992, A&A 256, 45
Mezger P.G., 1978, A&A 70, 565
Miller G.E., Scalo J.M., 1979, ApJS 41, 513
Osterbrock D.E., 1989, *Astrophysics of Gaseous Nebulae and Active Galactic Nuclei*, Oxford University Press
Pacholczyk A.G., 1970, *Radio Astrophysics*, W.H. Freeman and Company, San Francisco
Pérault M., Boulanger F., Puget J.L., Falgarone E., 1988, unpublished
Rice W., Boulanger F., Viallefond F., Soifer B.T., Freedman W.L., 1990, ApJ 358, 418
Ruzmaikin A., Sokoloff D., Shukurov A., 1988, Nature 336, 341
Savage B.D., Mathis S.M., 1979, A&AR 17, 73
Scalo J.M., 1986, Fundam. of Cosmic Research 11
Sofue Y., Fujimoto M., Wielebinski, R., 1986, A&AR 24, 459
van den Bergh S., Tammann G., 1991, A&AR 29, 363
Völk H.J., 1989, A&A 218, 67
Völk H.J., Zank L.A., Zank G.P, 1988, A&A 198, 274
Völk H.J., 1992, in *Particle Acceleration in Cosmic Plasma*, eds. G.P. Zank and T.K. Gaisser, AiP Conf. Proc. 264, AiP, New York, 199
Völk H.J., Xu C., 1994, Infrared Phys. Technol. 35, 527
Walterbos R.A.M., Kennicutt R.C.Jr., 1988, A&A 198, 61
Wunderlich E., Klein U., Wielebinski R., 1987, A&AS 69, 487
Xu C., 1990, ApJ 365, L47
Xu C., Buat V., 1995, A&A 293, L65
Xu C., De Zotti G., 1989, A&A 225, 12
Xu C., Lisenfeld U., Völk H.J., Wunderlich E., 1994, A&A 282, 19